\documentclass[superscriptaddress,prd,twocolumn,showpacs,preprintnumbers,amsmath,amssymb,bibnotes,altaffilletter]{revtex4} 

\usepackage{graphicx}
\usepackage{bm}
\usepackage{color}

\begin{document}

\title{A Search for Periodicities in the $^{\bm{8}}$B Solar Neutrino Flux\\
  Measured by the Sudbury Neutrino Observatory}

%
\newcommand{\ubc}{Department of Physics and Astronomy, University of
British Columbia, Vancouver, BC V6T 1Z1, Canada}
\newcommand{\bnl}{Chemistry Department, Brookhaven National
Laboratory,  Upton, NY 11973-5000}
\newcommand{\carleton}{Ottawa-Carleton Institute for Physics, 
Department of Physics, Carleton University, Ottawa, Ontario K1S 5B6, 
Canada}
\newcommand{\uog}{Physics Department, University of Guelph,
Guelph, Ontario N1G 2W1, Canada}
\newcommand{\lu}{Department of Physics and Astronomy, Laurentian
University, Sudbury, Ontario P3E 2C6, Canada}
\newcommand{\lbnl}{Institute for Nuclear and Particle Astrophysics and
Nuclear Science Division, Lawrence Berkeley National Laboratory, 
Berkeley, CA 94720}
\newcommand{\lbla}{ Lawrence Berkeley National Laboratory, Berkeley, CA}
\newcommand{\lanl}{Los Alamos National Laboratory, Los Alamos, NM 87545}
\newcommand{\llnl}{Lawrence Livermore National Laboratory, Livermore, CA}
\newcommand{\lanla}{Los Alamos National Laboratory, Los Alamos, NM}
\newcommand{\oxford}{Department of Physics, University of Oxford,
Denys Wilkinson Building, Keble Road, Oxford OX1 3RH, UK}
\newcommand{\penn}{Department of Physics and Astronomy, University of
Pennsylvania, Philadelphia, PA 19104-6396}
\newcommand{\queens}{Department of Physics, Queen's University,
Kingston, Ontario K7L 3N6, Canada}
\newcommand{\uw}{Center for Experimental Nuclear Physics and Astrophysics,
and Department of Physics, University of Washington, Seattle, WA 98195}
\newcommand{\uta}{Department of Physics, University of Texas at 
Austin, Austin, TX 78712-0264}
\newcommand{\triumf}{TRIUMF, 4004 Wesbrook Mall, Vancouver, BC V6T 2A3, Canada}
\newcommand{\ralimp}{Rutherford Appleton Laboratory, Chilton, Didcot 
OX11 0QX, UK}
\newcommand{\iusb}{Department of Physics and Astronomy, Indiana 
University, South Bend, IN}
\newcommand{\fnal}{Fermilab, Batavia, IL}
\newcommand{\uo}{Department of Physics and Astronomy, University of 
Oregon, Eugene, OR}
\newcommand{\rcnp}{Department of Physics, Osaka University, Osaka, Japan}
\newcommand{\snolab}{SNOLAB, INCO Creighton \#9 Mine, P.O. Box 159,
  Lively, Ontario   P3Y 1M3, Canada}
\newcommand{\slac}{Stanford Linear Accelerator Center, Menlo Park, CA}
\newcommand{\mac}{Department of Physics, McMaster University, Hamilton, ON}
\newcommand{\doe}{US Department of Energy, Germantown, MD}
\newcommand{\lund}{Lund University, Sweden}
\newcommand{\mpi}{Max-Planck-Institut for Nuclear Physics, Heidelberg, Germany}
\newcommand{\uom}{Ren\'{e} J.A. L\'{e}vesque Laboratory, 
Universit\'{e} de Montr\'{e}al, Montreal, PQ}
\newcommand{\cwru}{Department of Physics, Case Western Reserve 
University, Cleveland, OH}
\newcommand{\pnnl}{Pacific Northwest National Laboratory, Richland, WA}
\newcommand{\uc}{Department of Physics, University of Chicago, Chicago, IL}
\newcommand{\mitt}{Department of Physics, Massachusetts Institute of 
Technology, Cambridge, MA }
\newcommand{\ucsd}{Department of Physics, University of California at 
San Diego, La Jolla, CA }
\newcommand{	\lsu	}{Department of Physics and Astronomy, 
Louisiana State University, Baton Rouge, LA 70803}
\newcommand{\imp}{Imperial College, London SW7 2AZ, UK}
\newcommand{\uci}{Department of Physics, University of California, 
Irvine, CA 92717}
\newcommand{\ucia}{Department of Physics, University of California, Irvine, CA}
\newcommand{\suss}{Department of Physics and Astronomy, University of 
Sussex, Brighton  BN1 9QH, UK}
\newcommand{	\lifep	}{Laborat\'{o}rio de Instrumenta\c{c}\~{a}o e 
F\'{\i}sica Experimental de
Part\'{\i}culas, Av. Elias Garcia 14, 1$^{\circ}$, 1000-149 Lisboa, Portugal}
\newcommand{\hku}{Department of Physics, The University of Hong Kong, 
Hong Kong.}

\affiliation{	\ubc	}
\affiliation{	\bnl	}
\affiliation{	\carleton	}
\affiliation{	\uog	}
\affiliation{	\lu	}
\affiliation{	\lbnl	}
\affiliation{	\lanl	}
\affiliation{	\lifep	}
\affiliation{	\lsu	}
\affiliation{	\oxford	}
\affiliation{	\penn	}
\affiliation{	\queens	}
\affiliation{	\ralimp	}
\affiliation{	\uta	}
\affiliation{	\triumf	}
\affiliation{	\uw	}

\author{	B.~Aharmim	}			\affiliation{ 
	\lu	}
\author{	S.N.~Ahmed	}			\affiliation{ 
	\queens	}
\author{	A.E.~Anthony	}			\affiliation{ 
	\uta	}
\author{	E.W.~Beier	}			\affiliation{ 
	\penn	}
\author{	A.~Bellerive	}			\affiliation{ 
	\carleton	}
\author{	M.~Bergevin	}			\affiliation{ 
	\uog	}
\author{	S.D.~Biller	}			\affiliation{ 
	\oxford	}
\author{	M.G.~Boulay	}			\affiliation{ 
	\lanl	}
\author{	M.G.~Bowler	}			\affiliation{ 
	\oxford	}
\author{	Y.D.~Chan	}			\affiliation{ 
	\lbnl	}
% ...

\collaboration{SNO Collaboration}
\noaffiliation

\date{July 18, 2005; Revised August 4, 2005; to appear as Phys Rev D
  72, 052010}

\begin{abstract}
A search has been made for sinusoidal periodic variations in the $^8$B
solar neutrino flux using data collected by the Sudbury Neutrino
Observatory over a 4-year time interval.  The variation at a period of
one year is consistent with modulation of the $^8$B neutrino flux by the
Earth's orbital eccentricity.  No significant sinusoidal periodicities
are found with periods between 1 day and 10 years with either an
unbinned maximum likelihood analysis or a Lomb-Scargle periodogram
analysis.  The data are inconsistent with the hypothesis that the
results of the recent analysis by Sturrock {\em et al.}, based on
elastic scattering events in Super-Kamiokande, can be attributed to a
7\% sinusoidal modulation of the total $^8$B neutrino flux.
\end{abstract}


\maketitle


\section{Introduction}

There have been recent reports of periodic variations in the measured
solar neutrino fluxes~\cite{sturrock_sk1,sturrock_sk2,sturrock_sk3,
sturrock_gallex1,sturrock_gallex2,sturrock_gallex3,milsztajn,ranucci}.
Other analyses of these same data, including analyses by the
experimental collaborations themselves, have failed to find such
evidence ~\cite{sk, pandola}.  The reported periods have been claimed
to be related to the solar rotational period.  Particularly relevant
for this paper is a claimed 7\% amplitude modulation in
Super-Kamiokande's $^8$B neutrino flux at a frequency of
9.43~y$^{-1}$~\cite{sturrock_sk2,sturrock_sk3}.  Because solar
rotation should not produce variations in the solar nuclear fusion
rate, non-standard neutrino properties have been proposed as an
explanation.  For example, the coupling of a neutrino magnetic moment
to rotating magnetic fields inside the Sun might cause solar neutrinos
to transform into other flavors through a resonant spin flavor
precession mechanism~\cite{rsfp1,rsfp2,rsfp3}.  Periodicities in the
solar neutrino flux, if confirmed, could provide evidence for new
neutrino physics beyond the commonly accepted picture of
matter-enhanced oscillation of massive neutrinos.

This paper presents a search for periodicities in the data from the
Sudbury Neutrino Observatory (SNO).  SNO is a real-time, water
Cherenkov detector located in the Inco, Ltd. Creighton nickel mine
near Sudbury, Ontario, Canada~\cite{sno_nim}.  SNO observes
charged-current (CC) and neutral-current (NC) interactions of $^8$B
neutrinos on deuterons in 1~ktonne of D$_2$O, as well as
neutrino-electron elastic scattering (ES) interactions.  By comparing
the observed rates of CC, NC, and ES interactions, SNO has
demonstrated that a substantial fraction of $^8$B electron neutrinos
produced inside the Sun transform into other active neutrino
flavors~\cite{cc_prl,d2o_prl,dn_prl,salt_prl,nsp}.

SNO's combination of real-time detection, low backgrounds, and
sensitivity to different neutrino flavors give it unique capabilities in a
search for neutrino flux periodicities.  Chief among these is the
ability to do an {\em unbinned} analysis, in which the event times of
individual neutrino events are used as inputs to a maximum likelihood
fit.

This paper presents results from an unbinned maximum likelihood
analysis and a more traditional Lomb-Scargle periodogram analysis for
SNO's pure D$_2$O and salt phase data sets.  Previous analyses of data
from other experiments have used the Lomb-Scargle
periodogram~\cite{scargle} and binned maximum likelihood techniques to
search for periodicities in the solar neutrino data.  These data
generally consist of flux values measured in a number of time bins of
unequal size.  Because analyses of binned data can be sensitive to the
choice of binning, which can also produce aliasing effects, it is
desirable to avoid binning the data if possible.
Section~\ref{sec:data_sets} describes the data sets.
Section~\ref{sec:general_search} contains the results of a general
search for any periodicities with periods between 1 day and 10 years.
Section~\ref{sec:specific} presents limits on the amplitudes at two
specific frequencies: the 9.43~y$^{-1}$ modulation of the $^8$B
neutrino flux claimed by Sturrock {\em et
al.}~\cite{sturrock_sk2,sturrock_sk3}, and a yearly modulation due to
the Earth's orbital eccentricity.

\section{Description of the SNO data sets}
\label{sec:data_sets}

The data included in these analyses consist of the selected neutrino
events for the initial phase of SNO, in which the detector contained
pure D$_2$O~\cite{d2o_prl}, and for SNO's salt phase, in which 2
tonnes of NaCl were added to the D$_2$O to increase the neutron
detection efficiency for the NC reaction~\cite{nsp}.  Each data set is
divided into runs of varying length during which the detector was live
for solar neutrino events.  The D$_2$O data set consists of 559 runs
starting on November 2, 1999, and spans a calendar period of 572.2
days during which the total neutrino livetime was 312.9 days.  The
salt phase of SNO started on July 26, 2001, 59.7 calendar days after
the end of the pure D$_2$O phase of the experiment.  The salt data set
contains 1212 runs and spans a calendar period of 762.7 days during
which the total neutrino livetime was 398.6 days.  The intervals
between runs during which SNO was not recording solar neutrino events
correspond to run transitions, detector maintenance, calibration
activities, periods when the detector was off, etc.  Deadtime incurred
within a run, mostly due to spallation cuts that remove events
occurring within 20 seconds after a muon, can be neglected, since such
deadtime is incurred randomly at average intervals much shorter than
the periods of interest for this analysis.  This deadtime is 2.1\% for
the D$_2$O data set and 1.8\% for the salt data set.

The event selection for the data sets is similar to that in
\cite{d2o_prl} and \cite{nsp}.  Events were selected inside a
reconstructed fiducial volume of $R<550$~cm and above an effective
kinetic energy of $T_\text{eff}>5$~MeV (D$_2$O) or $T_\text{eff}>5.5$~MeV
(salt).  The salt data set contains 4722 events, as in \cite{nsp}.
During the salt analysis described in \cite{nsp} a background of
``event bursts'', consisting of two or three neutron-like events
occurring in a short time interval, was identified and removed with a
cut that eliminated any event occurring within 50~ms of an otherwise
acceptable candidate neutrino event.  The source of these 11 bursts is
not certain, but they may have been produced by atmospheric neutrino
interactions.  For this analysis a similar cut removing any event
occurring within 150~ms of another event was applied to the D$_2$O
data, reducing the number of selected events from 2928, as in
\cite{d2o_prl}, to 2924.  The timing window for the cut in the D$_2$O
data is longer than for the salt data to account for the longer
neutron capture time in pure D$_2$O.

An important element of a periodicity analysis is exact knowledge of
when each data-taking run began and ended.  These run boundaries
define the time exposure of the data set, which itself may induce
frequency components that could impact a periodicity analysis.  The
unbinned maximum likelihood analysis described below makes explicit
use of these run boundary times, and all Monte Carlo simulations are
generated using the exact run boundaries, even if the simulated data
are binned in a following analysis.  These precautions avoid {\em ad
hoc} assumptions about the distribution of the time exposure within
any time bin.  The measured time for each event was measured with a
global positioning system (GPS) clock to a precision of $\sim 100$~ns,
but rounded to 10~ms accuracy for the analysis.  The run boundary
times were determined from the times of the first and last events in
each run with a precision of $\sim 50$~ms.

\begin{figure}
\begin{center}
\includegraphics[width=3.6in]{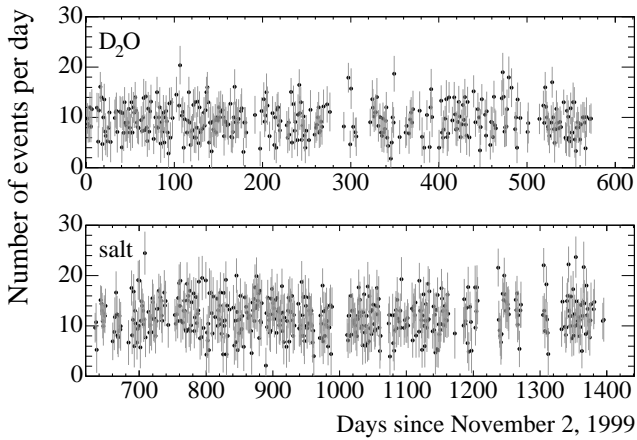}
\caption{\label{fig:1day-bins}
The livetime-corrected 1-day total rate of events as a function of time for the
D$_2$O and salt phases. The weighted mean rates are $9.35\pm0.17$ and
$11.85\pm0.17$ events/day, respectively.
}
\end{center}
\end{figure}

Figure~\ref{fig:1day-bins} 
displays the solar neutrino event rate in livetime corrected 1-day bins
over the total exposure time of both phases of SNO~\cite{sno_data}.  
The D$_2$O and salt data sets may be individually examined for
periodicities, or the combined data from both phases can be jointly
searched.  It should be noted that the relative amounts of CC, NC, and
ES events are different for the D$_2$O and salt data, with the salt
data set containing a much higher fraction of NC events.

Although SNO's data sets are dominated by solar neutrino
events, they also contain a small number of non-neutrino backgrounds,
primarily neutrons produced through photodisintegration of deuterons
by internal or external radioactivity.  The total estimated number of
background events is 123$^{+22}_{-17}$ for the D$_2$O data set
(4.2$^{+0.8}_{-0.6}$\% of the total rate), and 260$\pm 59$ events for
the salt data set (5.5$\pm 1.2$\% of the total rate).  Although the
background rate is not entirely constant, the backgrounds are small
and stable enough that they can be neglected in this analysis. 


\section{General Periodicity Search}
\label{sec:general_search}

Both an unbinned maximum likelihood analysis and a Lomb-Scargle
periodogram with 1-day binning were used to search SNO's data for
periodicities.  Results are presented below for the D$_2$O, salt, and
combined data from each method, along with evaluations of the
sensitivity of each method to sinusoidal variations of various periods and
amplitudes.  The periodicity searches were carried out over the sum of
CC, NC, and ES events.

Extensive use was made of Monte Carlo data sets to evaluate the
statistical significance of the results and the sensitivity of each
method.  To determine the statistical significance of any peak in the
frequency spectrum, 10,000 Monte Carlo data sets with events generated
randomly within the run boundaries for each phase were used, with mean
event rates in each phase matching those observed in SNO's data sets.
The number of events in each Monte Carlo data set was drawn from a
Poisson distribution with the same average rate as the data, and the
events were distributed uniformly within the run boundaries
~\cite{footnote1}.  These ``null-hypothesis'' Monte Carlo data sets
were used to determine the probability that a data set drawn from a
constant rate distribution would produce a false positive detection of
a periodicity.  To determine the sensitivity of an analysis to a real
periodicity, 1,000 Monte Carlo data sets were generated for each of
several combinations of frequencies and amplitudes, with the events
drawn from a time distribution of the form $1 + A \cos (2\pi f t)$.
The {\em sensitivity} for any frequency $f$ and amplitude $A$ is then
defined as the probability that the analysis will reject the null
hypothesis of a constant rate at the 99\% confidence level.

\subsection{Unbinned Maximum Likelihood Method}
\label{sec:ml}

The unbinned maximum likelihood method tests the hypothesis that the
observed events are drawn from a rate distribution given by
\begin{equation}
\phi(t) =  N (1 + A \cos (2\pi f t + \delta))
\label{eq:pdf}
\end{equation}
relative to the hypothesis that they are drawn from a constant rate
distribution ($A=0$).  $A$ is the fractional amplitude of the
periodic variation about the mean, $\delta$ is a phase offset, and
$N$ is a normalization constant for the rate.  Equation~\ref{eq:pdf}
serves as the probability density function (PDF) for the observed event times,
which are additionally constrained to occur only within run boundaries
(i.e., $\phi(t) = 0$ if $t$ is not between the start and end times of
any run).  

With $f$ fixed, the maximum of the extended likelihood
$L(t_k|N,A,\delta,f)$ as a function of the individual event times
$t_k$ is calculated for a data set as
\begin{equation}
\ln L(t_k|N,A,\delta,f) = -\sum_{j=1}^\text{runs} \int_{t_i^j}^{t_f^j}
\! \phi(t) dt + \sum_{k=1}^\text{events} 
\ln (\phi(t_k))
\label{eq:loglike}
\end{equation}
where the first term is a sum over all runs of an integral evaluated
between each run's start and stop times $t_i^j$ and $t_f^j$, and
accounts for Poisson fluctuations in the signal amplitude.  The second
term is a sum over the events in the data set, and $t_k$ is the time
of the $k$th event.  The log likelihood is maximized as a function of
$A$, $\delta$, and $N$ to yield $\ln L_\text{max}$, while $f$ is kept
fixed.  Then the constraint $A=0$ is imposed, removing the dependence
of $\phi(t)$ on both $A$ and $\delta$, and the log likelihood is
maximized over the remaining free parameter $N$ to yield $\ln
L_\text{max}(A=0)$.  By the likelihood ratio theorem~\cite{likelihood}
the difference $2S \equiv 2(\ln L_\text{max} - \ln L_\text{max}(A=0))$
will approximately have a $\chi^2$ distribution with two degrees of
freedom (since the choice $A=0$ also removes the dependence on the
phase $\delta$).  Thus $S$ will follow a simple exponential if the
true value of $A$ is zero.  Therefore, at any single frequency $f$,
the probability of observing $S>Z$ under the null hypothesis that
$\phi(t) = N$ is approximately $e^{-Z}$.  This null hypothesis test is
carried out for a large set of frequencies scanning the region of
interest.

Equation~\ref{eq:loglike} includes both a floating offset $N$ and an
amplitude $A$ as free parameters.  Allowing both of these parameters to
vary is necessary to deal with very low frequencies, for which $N$ and
$A$ become degenerate parameters.  Simply fixing $N$ to the mean rate,
as was done in \cite{sturrock_sk3}, will be prone to bias at the very
lowest frequencies, but gives virtually identical results to the
floating offset procedure when the length of the data set is longer
than the period $T = 1/f$, since in this case enough cycles
are sampled to break the degeneracy between $N$ and $A$.

Equations~\ref{eq:pdf} and \ref{eq:loglike} are adequate to test for
periodicity in a single data set, but for a combined analysis of SNO's
D$_2$O and salt data sets, account must be taken of the differing mean
rates owing to different detection efficiencies and energy thresholds
in the two phases.  This can be done by generalizing $\phi(t)$
to:
\begin{eqnarray}
\phi(t) & = & N_\text{d2o} (1 + A \cos (2\pi f t + \delta)), {\rm ~if~} t \in
    {\rm ~D}_2{\rm O~run}
\nonumber \\
\phi(t) & = & N_\text{salt} (1 + A \cos (2\pi f t + \delta)), {\rm ~if~} t
    \in {\rm ~salt~run} \nonumber 
\label{eq:combined_pdfs}
\end{eqnarray}
This PDF allows different normalization constants for the two data
sets, while retaining the assumption that the flux variation
has the same fractional amplitude in both the D$_2$O and salt
data. 

\subsubsection{Results for the SNO data sets}
\label{sec:ml_results}

\begin{figure}
\begin{center}
\includegraphics[width=3.6in]{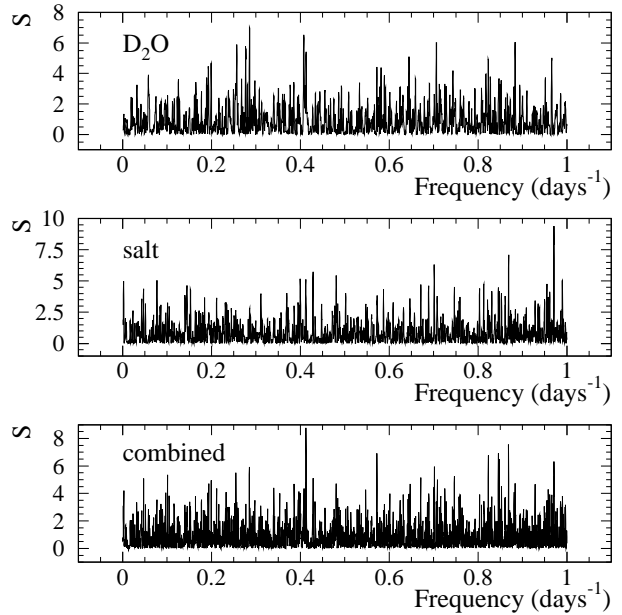}
\caption{\label{fig:ml_results}
Log likelihood difference ($S$) as a function of frequency for the unbinned
maximum likelihood method for SNO's D$_2$O, salt, and combined data sets.
}
\end{center}
\end{figure}

Figure~\ref{fig:ml_results} shows $S$ as a function of frequency for
the D$_2$O, salt, and combined data sets at 3650 frequencies with
periods ranging from 10 years down to 1 day, with a sampling interval
of $\Delta f = 1/(3650~{\rm days})$.  This corresponds to an
oversampling of the number of independent Fourier frequencies for
continuous data by a factor of approximately 5-6 for the separate D$_2$O
and salt data sets, and a factor of 2.6 for the combined data set.
The maximum value of $S$ for the D$_2$O data set is $S=7.1$ at a
period of 3.50 days ($f = 0.296$~days$^{-1}$).  The largest peak found
in the salt data has a height of $S=9.4$ at a period of 1.03 days
($f=0.971 $~days$^{-1}$), while the combined data set has its largest
peak of $S=8.8$ at a period of 2.40 days ($f=0.417$~days$^{-1}$).

\begin{figure}
\begin{center}
\includegraphics[width=3.6in]{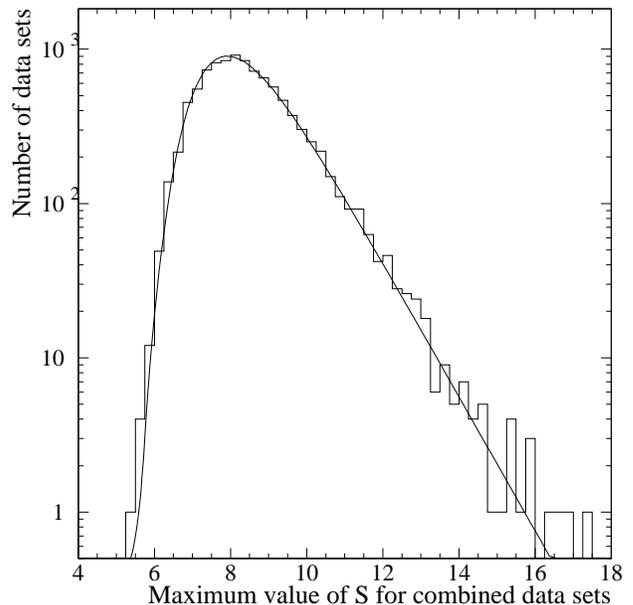}
\caption{\label{fig:maxpeak_dist} Distribution of the maximum value of $S$
for 10,000 Monte Carlo data sets produced with $A=0$ for a combined
D$_2$O+salt unbinned maximum likelihood analysis.  A fit of the
distribution to Equation~\ref{eq:indepprobs} is shown.  }
\end{center}
\end{figure}

Figure~\ref{fig:maxpeak_dist} shows the distribution of maximum peak
heights for 10,000 Monte Carlo data sets generated with no periodicity
and analyzed identically to SNO's combined data set.
Distributions for the D$_2$O and salt Monte Carlo data sets analyzed
individually look similar.  Of the 10,000 simulated data sets, 35\%
yielded at least one peak with $S>8.8$, exceeding the largest peak
seen in SNO's combined data set.  For the D$_2$O Monte Carlo data
sets, 72\% had a peak larger than the observed largest peak of
$S=7.1$, while 14\% of the salt Monte Carlo data sets yielded a peak larger
than the $S=9.4$ peak seen in the data.  Therefore, none of the
observed peaks are statistically significant.

Under the null hypothesis of no time variability, the probability of
any individual frequency having $S$ smaller than some threshold $Z$ is
approximately $1-e^{-Z}$.  If all 3650 scanned frequencies were statistically
independent, the probability that all peaks would be smaller than $Z$
would be $(1-e^{-Z})^{3650}$.  However, the 3650 scanned frequencies
are not strictly independent, since a finite data set has limited
frequency resolution, and neighboring frequencies are correlated.
If $F$ is the effective number of independent frequencies, then the
probability distribution for the height $Z$ of the largest peak 
approximately follows
\begin{equation}
P(Z)~dZ \propto e^{-Z} (1-e^{-Z})^{F-1}~dZ
\label{eq:indepprobs}
\end{equation}
The effective number of independent frequencies increases with the
length of the data set and number of detected events.  Fitting the
Monte Carlo distributions for $Z$ to this equation yields $F = 1422
\pm 17$ for the D$_2$O, $F=1696 \pm 26$ for the salt, and $F = 2739
\pm 30$ for the combined data set.  Figure~\ref{fig:maxpeak_dist}
shows this fit for the combined analysis.  These values are consistent
with expectations based on the oversampling factors described in
Section~\ref{sec:ml_results}~\cite{orford}.  Although
Equation~\ref{eq:indepprobs} appears to model $P(Z)$ well, quoted
significance levels are always determined directly from the Monte
Carlo distributions and not from the analytic formula.  To ensure that
no significant peaks were missed, the combined analysis of the actual
data (but not the Monte Carlo data sets) was repeated with the
sampling increased by a factor of five.  No new peaks were found.

\subsubsection{Sensitivity to sinusoidal periodicities}
\label{sec:ml_sensitivity}

Distributions of the maximum peak height for Monte Carlo data sets,
such as in Figure~\ref{fig:maxpeak_dist}, readily yield the threshold
$\zeta$ for which 99\% of Monte Carlo data sets generated without
periodicity would yield a maximum peak height of $\zeta$ or less.
This threshold defines the peak height at which the null hypothesis of
no time variation is rejected at the 99\% confidence level, and equals
12.10, 12.20, and 12.65 for the D$_2$O, salt, and combined data sets
respectively.

\begin{figure}
\begin{center}
\includegraphics[width=3.6in]{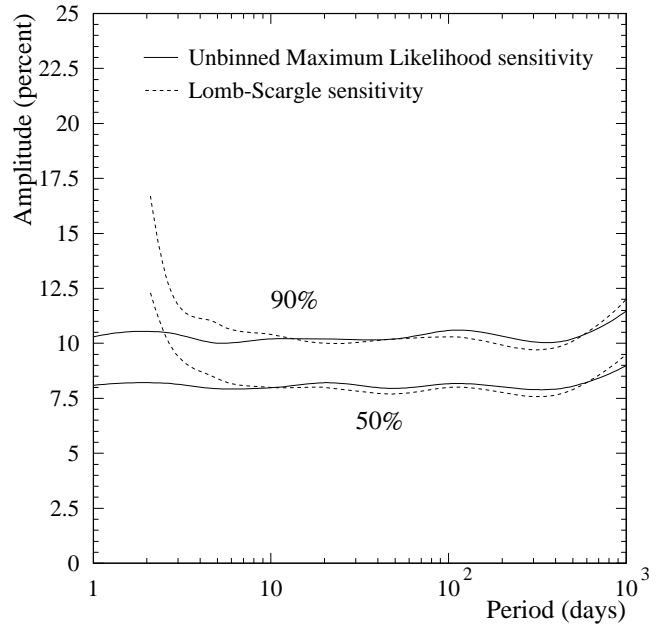}
\caption{\label{fig:sensitivity} Sensitivity contours for the unbinned
maximum likelihood and Lomb-Scargle methods.  The contours indicate
the amplitude as a function of period at which the analysis method
has a 50\% (90\%) chance of discovering a sinusoidal variation of
unknown period at the 99\% confidence level.  For longer periods both
methods are equally sensitive.  
Due to its lack of binning, the maximum likelihood method has
sensitivity to much smaller periods than the Lomb-Scargle method.
}
\end{center}
\end{figure}
Monte Carlo data sets drawn from rate distributions with sinusoidal
periodicities of various periods and amplitudes were analyzed to
determine the probability of rejecting the null hypothesis at the 99\%
C.L.

Figure~\ref{fig:sensitivity} shows the
amplitudes as a function of period at which the method has a 50\%
(90\%) probability of rejecting the null hypothesis, for simulations
of SNO's combined data set.  While the sensitivity varies as a
function of period, a signal must have an amplitude of approximately
8\% to be discovered 50\% of the time.

\subsection{The Lomb-Scargle periodogram}

The Lomb-Scargle periodogram is a method for searching unevenly sampled
data for sinusoidal periodicities~\cite{scargle} and 
provides an alternative to the unbinned maximum likelihood technique
described above.

The Lomb-Scargle power $P(f)$ at  frequency $f$ is calculated
from the measured flux values $y(t_i)$ in $N$ independent time bins as:
\begin{widetext}
\begin{equation}
P(f)=  \frac{1}{2\sigma^2} \left(
\frac{\left[\sum_{i=1}^{N} w_i (y(t_i) - \bar{y}) \cos (2\pi f(t_i-\tau))
\right]^2}
{\sum_{i=1}^N w_i \cos^2 (2\pi f(t_i-\tau))} \\
+ \frac{\left[\sum_{i=1}^{N} w_i (y(t_i) - \bar{y}) \sin (2\pi f(t_i-\tau))
\right]^2}
{\sum_{i=1}^N w_i \sin^2 (2\pi f(t_i-\tau))}
\right)\\
\label{eq:ls1}
\end{equation}
\end{widetext}
where the phase factor $\tau$ satisfies:
\begin{displaymath}
{\rm tan} (4\pi f\tau) = \frac{\sum_{i=1}^N w_i \sin (4\pi f
  t_i)}{\sum_{i=1}^N w_i \cos (4\pi f t_i)} 
\label{eq:ls2}
\end{displaymath}
Each bin is weighted in proportion to the inverse of its squared
uncertainty divided by the average value of the inverse of the squared
uncertainty $\langle 1/\sigma_i^2 \rangle$ (so $w_i = 1/\sigma_i^2 /
\langle 1/\sigma_i^2 \rangle$), as in \cite{sturrock_sk3}.  In
Equation~\ref{eq:ls1} $t_i$ is the livetime-weighted mean time for the
$i$th bin, and $\bar{y}$ and $\sigma^2$ are the weighted mean and
weighted variance of the data for all the bins calculated with the
weighting factors $w_i$.

Like the maximum likelihood method, the power $P$ in the Lomb-Scargle
periodogram at any single  frequency $f$ is expected to
approximately follow an
exponential distribution $e^{-P}$ if the data set is drawn from a
constant rate distribution.  The same methods of evaluating the
significance of the largest peak and the sensitivity of the method to
periodic signals can be employed, making use of large numbers of Monte
Carlo data sets.

In \cite{sk} the Super-Kamiokande collaboration used an unweighted
Lomb-Scargle periodogram ($w_i \equiv 1)$ to search its data set for
periodicities, a choice that was criticized in \cite{sturrock_sk3}.
The analysis presented here used the weighted Lomb-Scargle
periodogram.

For the Lomb-Scargle method SNO's recorded events were binned in 1-day
intervals (see Figure~\ref{fig:1day-bins}), 
and the livetime, the livetime-weighted mean time $t_i$, and the event
rate $y(t_i)$ were calculated for each bin.  To prevent biases
stemming from the assumption of Gaussian statistics,
any bin in which fewer than five events would be
expected based upon that bin's livetime and the mean event rate was
combined with the following bin(s) so that the expected number of events
in all bins was greater than five.  
The uncertainty $\sigma_i$ on the rate in each bin was taken to be the
square root of the expected number of events in that bin for a
constant rate.  This calculation of the uncertainty is appropriate if
one views the Lomb-Scargle method as a null hypothesis test of the
no-periodicity hypothesis; however, using the observed number of
events instead to calculate $\sigma_i$ does not change the conclusions
of this study.

When doing a combined D$_2$O + salt analysis one must account for the
different mean event rates in the two phases.  In the Lomb-Scargle
analysis this was accomplished by scaling the rates and uncertainties
on the salt data bins by the ratio of the weighted mean D$_2$O rate 
to the
weighted mean salt rate.  

\subsubsection{Results for the SNO data sets}

\begin{figure}
\begin{center}
\includegraphics[width=3.6in]{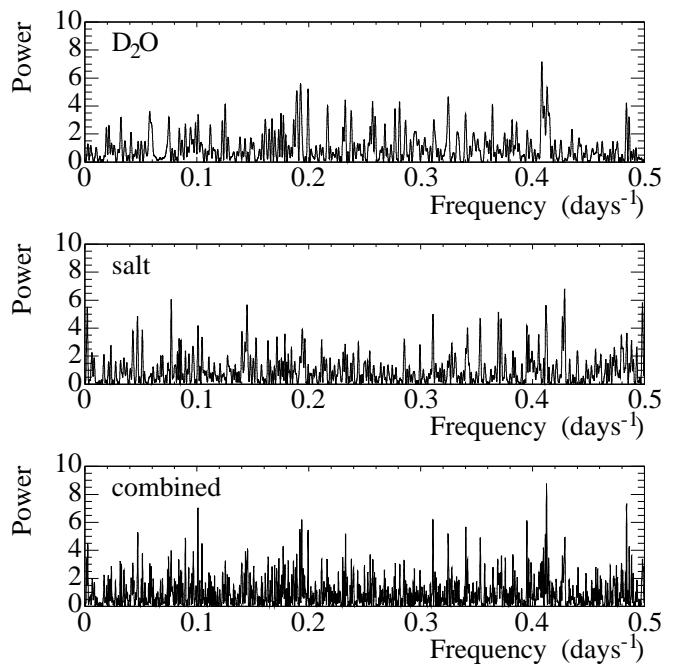}
\caption{\label{fig:ls_results}
Lomb-Scargle periodograms for SNO's D$_2$O, salt, and combined data
sets, with 1-day binning. 
}
\end{center}
\end{figure}

Figure~\ref{fig:ls_results} shows the Lomb-Scargle periodograms for
the D$_2$O, salt, and combined data sets.  A total of 7300 frequencies
were tested with periods ranging from 10 years to 2 days, with a
sampling step of $\Delta f = 1/(14600~{\rm days})$~\cite{footnote2}.
Because the data were binned in one-day intervals, the analysis was
restricted to frequencies less than 0.5~days$^{-1}$ to avoid potential
binning effects.  The maximum peak height for the D$_2$O data set is
$P=7.1$ at a period of 2.45 days ($f=0.408$~days$^{-1}$).  The largest
salt peak has a height of $P=6.8$ at a period of 2.33 days
($f=0.429$~days$^{-1}$), while the combined data set has its largest
peak of $P=8.7$ at a period of 2.42~days ($f=0.413$~days$^{-1}$).

The probability of observing a larger peak than that actually seen in
the Lomb-Scargle periodogram, if the rate were constant, was estimated
using the previously described 10,000 Monte Carlo data sets having no
periodicity.  Under the null hypothesis of no time variability, the
probability of getting a peak larger than the biggest peak seen in the
Lomb-Scargle periodogram is 46\% for the D$_2$O data set, 65\% for
the salt data set, and 27\% for the combined data sets.  As with the
unbinned maximum likelihood method, no evidence for time variability
is seen.

\subsubsection{Sensitivity to sinusoidal periodicities}
\label{sec:ls_sensitivity}

\begin{figure}
\begin{center}
\includegraphics[width=3.6in]{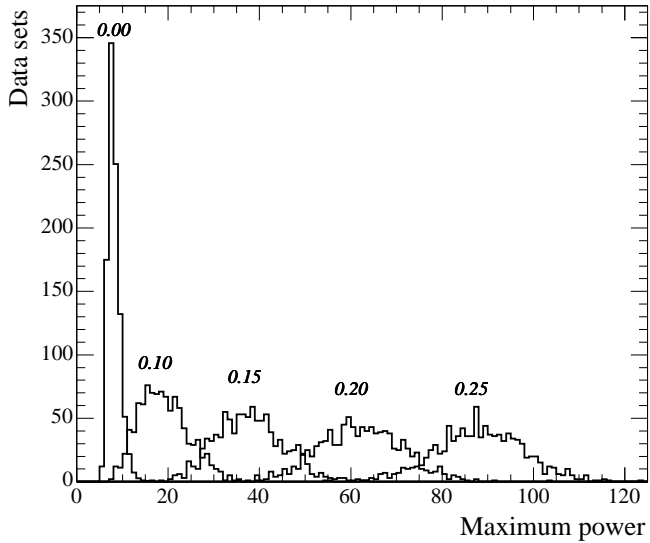}
\caption{\label{fig:ls_maxpeaks} 
Distribution of maximum Lomb-Scargle powers for 1,000 Monte Carlo data sets
produced with amplitudes $A=0.0,0.10,0.15,0.20,0.25$ and a period of 20 days 
for the combined D$_2$O+salt analysis.
}
\end{center}
\end{figure}

Monte Carlo data sets
generated with sinusoidal periodicities were used to estimate the
sensitivity of the Lomb-Scargle method to signals of
various periods and amplitudes.  Figure~\ref{fig:sensitivity} shows
the amplitudes as a function of frequency at which the analysis would
detect the signal 50\% or 90\% of the time.  In each case the
signal is considered to be detected if the Lomb-Scargle method rejects
the null hypothesis of a constant rate at the 99\% confidence level.
The threshold for rejecting the null hypothesis at the 99\% C.L. is
$P=11.15$ for the D$_2$O data, $P=11.43$ for the salt data, and
$P=12.24$ for the combined analysis. Figure~\ref{fig:ls_maxpeaks} shows
example maximum power distributions for a 20-day period with amplitudes
of 0, 10, 15, 20, and 25\% for the combined analysis.

\subsubsection{Systematic checks}

Many checks of the Lomb-Scargle periodogram were made to verify that
the results are robust. In particular, all data and Monte Carlo
results were recomputed for (a) a range of bin sizes, from 1-day to
5-days in fractional day steps, (b) a range of starting times of the
first bin in fractional day steps, and (c) different values of the
frequency sampling step.  There was no evidence for time
variability under any of these scenarios.

\section{Limits at Specific Frequencies of Interest}
\label{sec:specific}

The sensitivity calculations in Sections~\ref{sec:ml_sensitivity} and
\ref{sec:ls_sensitivity} are appropriate when the frequency of the
signal is not known {\em a priori}, and could occur anywhere in the
frequency search band.  The threshold for claiming a detection at the
99\% C.L. must accordingly be set relatively high to reduce the false
alarm probability, which was found from Monte Carlo simulations but is
approximately given by integrating Equation~\ref{eq:indepprobs} above
the detection threshold, to $<1\%$.  However, if the frequency of
interest is specified {\em a priori}, then a more restrictive and
sensitive test can be done using the fitted amplitude at that
frequency.  Two particular frequencies of interest are the 7\%
variation in the Super-Kamiokande data at a frequency of
$9.43$~y$^{-1}$ claimed by Sturrock {\em et al.}~\cite{sturrock_sk3}, and
the annual modulation of the neutrino flux by the Earth's orbital
eccentricity.

\subsection{Test at ${\bm{f = 9.43}}$~y$^{\bm{-1}}$}

Sturrock {\em et al.} have claimed evidence for a periodicity in
Super-Kamiokande's neutrino data at a frequency of $9.43 \pm
0.05$~y$^{-1}$ (0.0258~days$^{-1}$) with an amplitude of
7\%~\cite{sturrock_sk3}.  Examination of SNO's unbinned maximum
likelihood results in the interval from 9.33-9.53~y$^{-1}$ yielded no
value larger than $S=1.1$ in either the D$_2$O, the salt, or the
combined data sets~\cite{sk_comparison}.  The best-fit amplitude for
the combined data set inside this frequency interval is $(1.3 \pm
1.6)\%$.  This disagrees with a 7\% amplitude periodicity in the $^8$B
neutrino flux by 3.6 sigma.  It must be remarked that SNO's limit
applies to a modulation of the summed rates of CC, ES, and NC events
above their respective energy thresholds, whereas the reported 7\%
periodicity in the Super-Kamiokande data is a modulation of the
elastic scattering rate from $^8$B neutrinos above a total electron
energy threshold of 5~MeV.  The best-fit amplitudes for the D$_2$O and
salt data sets are $(3.8 \pm 2.6)\%$ and $(0.3 \pm 2.3)\%$
respectively.

\subsection{Eccentricity Result}

\begin{figure}
\begin{center}
\includegraphics[width=3.6in]{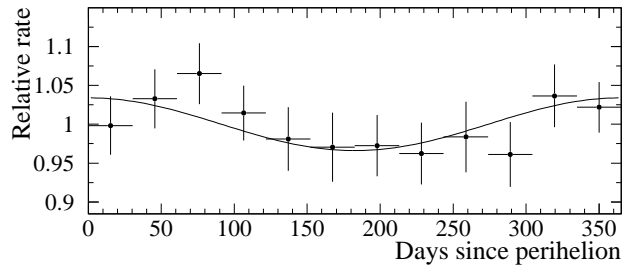}
\caption{\label{fig:eccentricity}
Relative event rate as a function of days since perihelion, normalized
to the mean rate.
In this plot the mean D$_2$O event rate has been scaled to match the
mean salt phase event rate, and the two data sets have been combined.
The curve represents the expected variation due to the eccentricity of
the Earth's orbit.
}
\end{center}
\end{figure}

The Earth's orbital eccentricity is expected to produce a rate
variation proportional, in excellent approximation, to $(1 + \epsilon
\cos (2\pi f(t-t_0)))^2$, where $\epsilon = 0.0167$ is the eccentricity
of the orbit, $f = 1/(365.25~{\rm days})$ is the Earth's orbital
frequency, and $t_0$ is the time of perihelion.  Maximum sensitivity
to this effect is obtained if $t_0$ and $f$ are fixed to their known
values and the combined data sets are fit for $\epsilon$ only.  This
has been implemented using the unbinned maximum likelihood technique.
The best-fit eccentricity is $\epsilon = 0.0143 \pm 0.0086$, in good
agreement with the expected value.  The difference in the log
likelihoods for the best fit compared to $\epsilon \equiv 0$ is 1.394.
The probability of obtaining a larger value of the log likelihood
difference if $\epsilon = 0$ is 9.5\%.  Figure~\ref{fig:eccentricity}
displays the relative event rate for the combined data as a function
of the time since perihelion.

\section{Conclusions}
\label{sec:conclusions}

Data from SNO's D$_2$O and salt phases have been examined for time
periodicities using an unbinned maximum likelihood method and the
Lomb-Scargle periodogram.  No evidence for any sinusoidal variation is
seen in either data set or in a combined analysis of the two data
sets.  This general search for sinusoidal variations with periods
between 1 day and 10 years has significant sensitivity to
periodicities with amplitudes larger than $\sim 8\%$.  The
best-fit amplitude for a sinusoidal variation in the total $^8$B neutrino flux
at a frequency of 9.43~y$^{-1}$ is $(1.3 \pm 1.6)$\%, which is
inconsistent with the hypothesis that the results of the recent
analysis by Sturrock {\em et al.}~\cite{sturrock_sk3}, based on elastic
scattering events in Super-Kamiokande, can be attributed to a 7\%
modulation of the $^8$B neutrino flux.  A fit for the eccentricity of the
Earth's orbit from the modulation at a period of one year yields
$\epsilon = 0.0143 \pm 0.0086$, in good agreement with the known value
of 0.0167.


\section*{ACKNOWLEDGMENTS}
This research was supported by: Canada: Natural Sciences and
Engineering Research Council, Industry Canada, National Research
Council, Northern Ontario Heritage Fund, Atomic Energy of Canada,
Ltd., Ontario Power Generation, High Performance Computing Virtual
Laboratory, Canada Foundation for Innovation; US: Dept.\ of Energy,
National Energy Research Scientific Computing Center; UK: Particle
Physics and Astronomy Research Council.  This research has been
enabled by the use of WestGrid computing resources, which are funded
in part by the Canada Foundation for Innovation, Alberta Innovation
and Science, BC Advanced Education, and the participating research
institutions. WestGrid equipment is provided by IBM, Hewlett Packard
and SGI.  We thank the SNO technical staff for their strong
contributions.  We thank Inco, Ltd. for hosting this project.



\begin{thebibliography}{100}
\bibitem{footnote1} Ideally, this sample would be drawn from a
Poisson distribution based on the ``true'' source rate, which is
unknown. However, in practice, this approximation has negligible
effect on the results presented here.
\bibitem{footnote2} The calculational ease of the Lomb-Scargle
analysis permitted a smaller sampling step than was used for the
unbinned maximum likelihood analysis, which was CPU-intensive.  We
have verified that the sampling steps used for both analyses were
adequate.
\end{thebibliography}
\end{document}